# Impact of roughness on heat conduction involving nanocontacts

**Eloïse Guen[1], Pierre-Olivier Chapuis[1], Nupinder Jeet Kaur[2,3], Petr Klapetek[2,3], and Séverine Gomès[1]***

[1]*Centre d'Energétique et de Thermique de Lyon (CETHIL), UMR CNRS 5008, INSA Lyon, UCBL, Université de Lyon, Villeurbanne, France*
[2]*Czech Metrology Institute (CMI), Okružní 31, 638 00 Brno, Czech Republic*
[3]*CEITEC, Brno University of Technology, Purkyňova 123, 612 00 Brno, Czech Republic*

**severine.gomes@insa-lyon.fr*

**The impact of surface roughness on conductive heat transfer across nanoscale contacts is investigated by means of scanning thermal microscopy. Silicon surfaces with out-of-plane rms roughness of ~0, 0.5, 4, 7 and 11 nm are scanned both under air and vacuum conditions. Three types of resistive SThM probes spanning curvature radii over orders of magnitude are used. A correlation between thermal conductance and adhesion force is highlighted. In comparison with a flat surface, the contact thermal conductance can decrease as much as 90% for a microprobe and by about 50% for probes with curvature radius lower than 50 nm. The effects of multi-contact and ballistic heat conduction are discussed. Limits of contact techniques for thermal conductivity characterization are also discussed.**

Studying heat transfer across solid contacts with nano-scaled imperfections is crucial for many industrial applications involving micro/nano-components as in electronics [1, 2]. Nanoscale roughness depends on fabrication processes and its impact on the thermal transport across interfaces can even dictate the overall thermal resistance in nanosystems [3]. From a fundamental point of view, understanding thermal transport between two solids is important when the characteristic dimensions in the zone of thermal contact become comparable to key length scales such as the mean free paths and the wavelengths of the energy carriers or the atomic distances of the materials in contact [4-7].

Measurements are usually performed over areas with transverse characteristic sizes larger than the micrometer [3], a scale where many nano-contacts may be present. There is hope that novel spatially-resolved nanocharacterization methods based on scanning probe measurements (SPM) can allow for more systematic studies of the single contact (constriction), or at least of regions involving a limited number of contacts. Scanning thermal microscopy (SThM), i.e. SPM with a thermal sensor on the tip, allows for coupled nanoscale analyses of heat transfer and contact mechanics [7]. A previous SThM study of polished nanoscale contacts [4] suggested that roughness down to the atomic scale is important, underlining possible effects of thermal quantization across individual atomic-scale contacts. It is clear that surface roughness alters the mechanical contact at many different scales, inducing discontinuous and reduced multi-contacts, which in the majority of cases decreases the total thermal transfer [8]. Roughness also impacts the shape of the humidity-induced water meniscus located around the mechanical contacts, which can impact the heat transfer at the probe-sample contact [9].

Here, thermal conductances between SThM tips of varying curvature radii and well-characterized rough silicon surfaces are determined, allowing to probe different contact scales. The total thermal contact radius and that due to the actual mechanical contact are discriminated while thermal results are correlated to adhesion forces. Thermal conductance is found to relate to the apparent contact radius at zero-force. It is also shown that the effect of roughness is comparable to that of an additional insulating layer, which can in particular be detrimental to SPM thermal measurements.

Three commercial resistive SThM probes were used: (1) the Wollaston wire probe [10] involving a 5-µm diameter wire with asperities at the apex, (2) the Pd probe [11] of Kelvin NanoTechnology where a palladium strip of curvature radius close to 100 nm

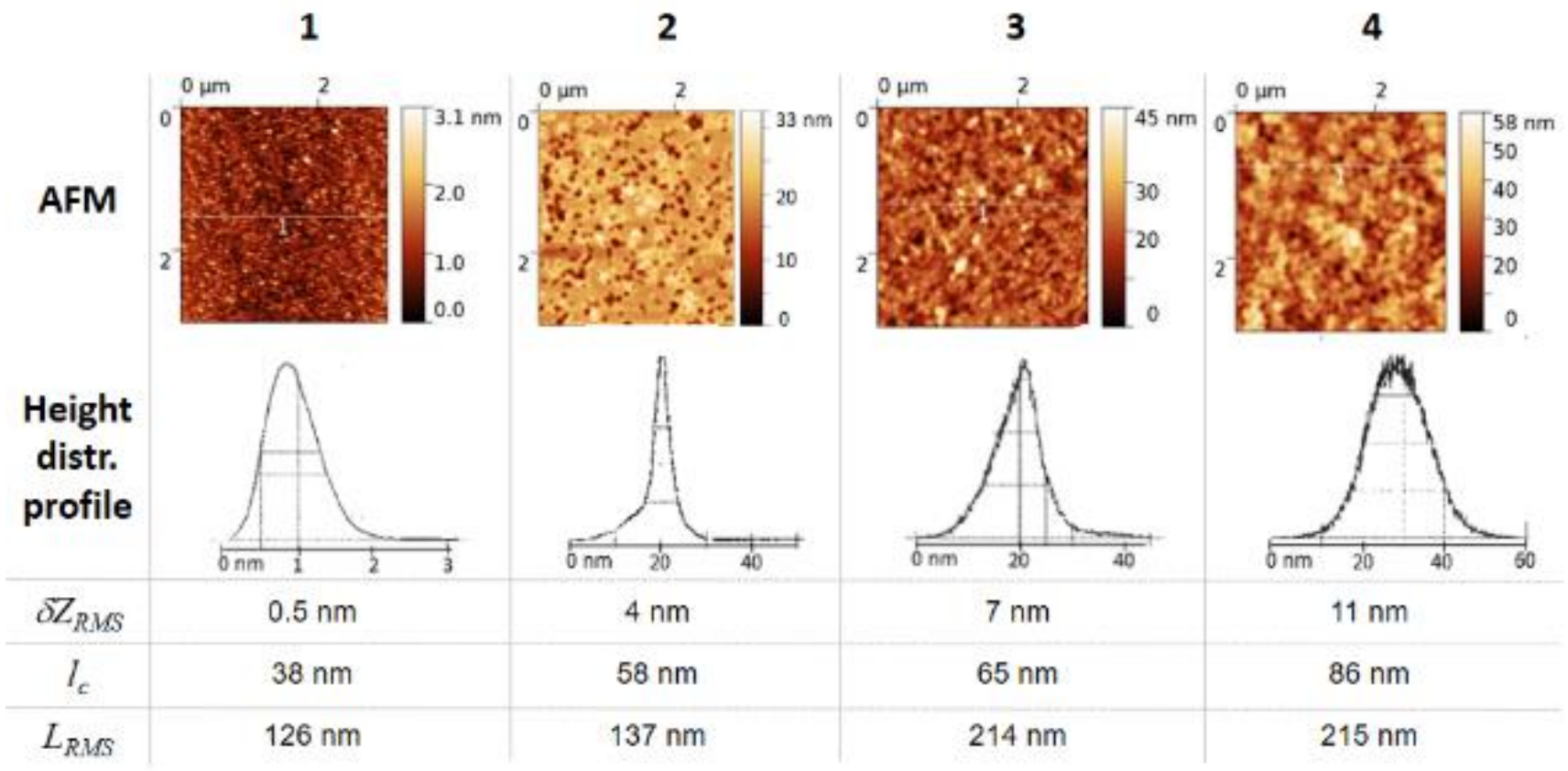

**Fig. 1. Topographic images obtained from AFM, distributions of the associated heights and roughness parameters (δ$Z_{RMS}$, $l_c$ and $L_{RMS}$) for the four rough samples**

is located at the apex, and the doped silicon (DS) probe [12] (AN300 thermal lever from Anasys Instruments) involving a ~10 nm curvature radius in silicon (see details in Suppl. Mat.). In the so-called 'active mode', the probe resistive element is self-heated by Joule effect in dc regime using a constant electrical current. After calibration (details in Suppl. Mat.), control units based on Wheatstone bridges were used to monitor their mean temperature rise $\bar{\theta}_p$ and the electrical power $P_{el}$ dissipated in the probe.

The samples consist of four silicon surfaces that have differing roughness parameters, prepared by anodic oxidation [13] and characterized by atomic force microscopy (AFM) (Fig. 1) by means of their root mean square roughness δ$Z_{RMS}$, transverse correlation length $l_c$ and mean peak-to-peak distance ($L_{RMS}$). All these parameters allow each sample to be accurately characterized in both the perpendicular and parallel directions to the sample. One can note a correlation between the trends of δ$Z_{RMS}$ and $l_c$. In addition, an untreated sample of smooth silicon substrate (δ$Z_{RMS}$ < 1 nm) from the same batch is used as a reference.

To assess the impact of surface roughness, measurements based on (i) AFM vertical approach curves and (ii) images obtained by *xy* scanning were both made in ambient air and in primary vacuum (pressure $P$ ~ 0.28 mbar), where the air contribution to the tip-sample transfer is eliminated. Results are provided as thermal conductances (see Suppl. Mat. for details on protocols). Fig. 2 reports on the decrease of thermal conductance (a) due to the global thermal transfer ($\Delta\bar{G}_{\text{global}}$) and (b) due to the tip jump to contact ($\Delta G^{a}_{mecha}$), for the three probes and both types of experimental conditions. Mechanical contacts forms after the jump, possibly with the water meniscus. We note strong differences between the behaviours with the different probes. When sample roughness increases, $\Delta\bar{G}_{\text{global}}$ decreases by 30% for the Wollaston probe (Fig. 2(a)A) and by about 10% for the Pd probe (Fig. 2(a)B). For the DS probe, $\Delta\bar{G}_{\text{global}}$ remains constant (Fig. 2(a)C). The observed conductance decreases are signatures of the decrease in heat conduction through the mechanical contact, as the air heat transfer taking place over a ~ micrometric zone is not expected to vary much when roughening the surface. For the largest probe (Wollaston, Fig. 2(a)), it is found that the heat conduction by mechanical contact (solid-solid and water meniscus) on a flat surface represents 20% of the overall transfer. This thermal transfer can almost be suppressed by roughening the surface (decrease by 95%) (Fig. 2A(b)). The overall decrease can be larger than 20%, so heat transfer through air is also slightly reduced, probably due to an effective tip-sample distance larger in the rough case. For the roughest sample, the mechanical contact accounts for only 2% of the overall heat transfer. In contrast, thermal transfer across mechanical contact accounts for less than 1% of the overall transfer on a flat surface for the smallest tip apex (DS probe, Figs. 3C). Although this transfer decreases with increased roughness, it has no visible effect on the measured overall transfer. Finally, the Pd probe, which has intermediate dimensions, exhibits an intermediate behavior (Figs. 3B(a)-(b)). On a flat surface, about 11% of the global transfer is made through mechanical contact. With the increase in roughness this transfer decreases by up to 30%, resulting in a 5-10% decrease on the overall signal measured. These results on thermal transfer across the first contact during an approach curve of the SThM tip on the sample show that surface roughness results in a decrease in the heat transfer across the contact, presuming a decrease in the probe-sample contact area. Analysis of thermal images of samples leads to the same conclusion (see Suppl. Mat. for images).

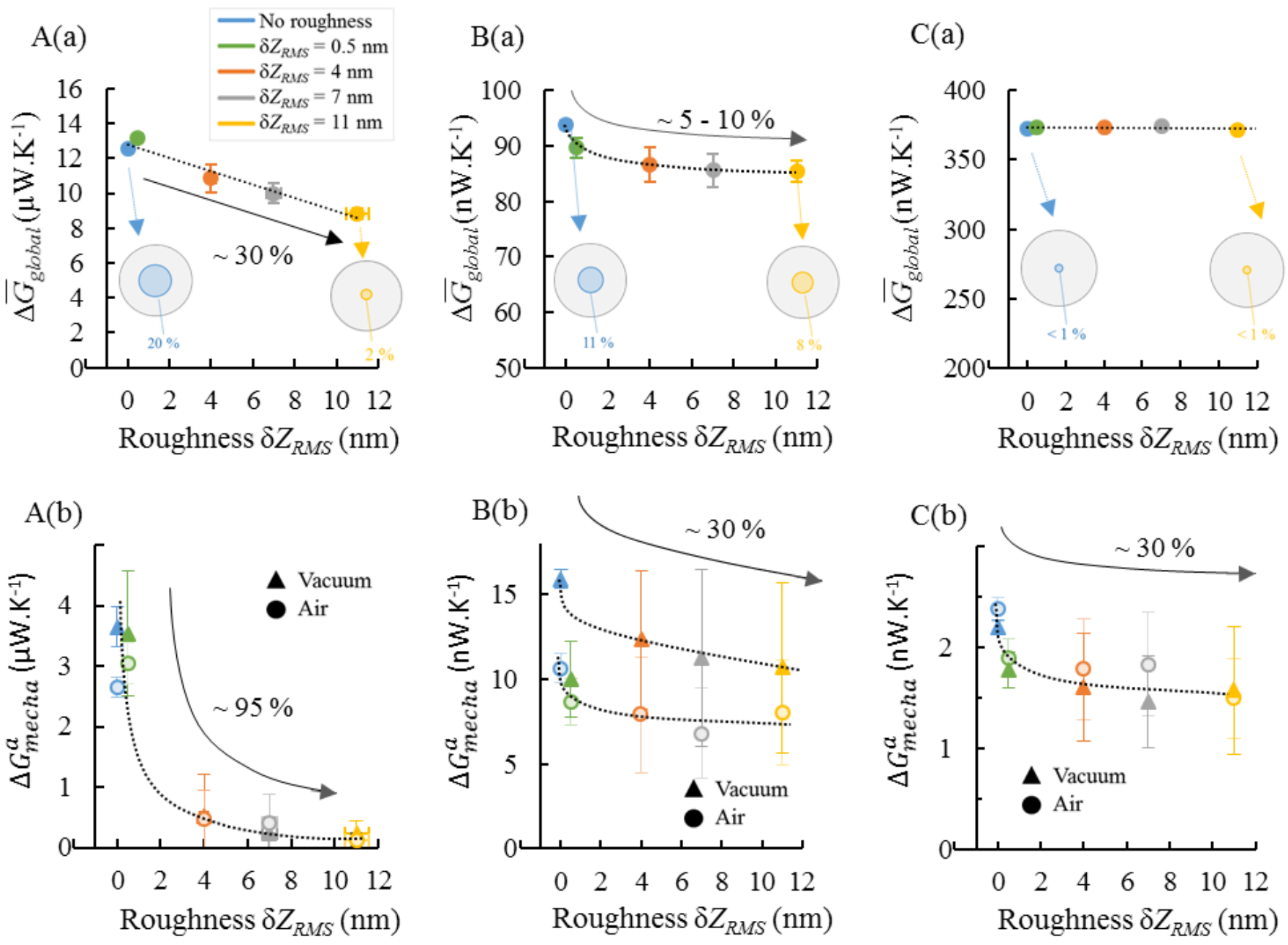

**Fig. 2. Measured global (a) and mechanical contact-related (b) thermal conductances according to the Si roughness, for the Wollaston (A), Pd (B) and doped Si (C) probes. Figs. (a) refer only to air measurements. Inset schematics represent the percentage of the mechanical contact heat transfer in the global heat transfer between probe and sample. For vacuum measurements $\Delta\bar{G}_{global} = \Delta G^a_{mecha}$ . Error bars represent dispersion of the measurements.**

Measurements of the adhesion forces [14] performed with the three probes on the rough samples are consistent with this observation. Fig. 3 shows the thermal conductance $\Delta G^a_{mecha}$ as a function of the average value of the adhesion force ($F_{ad}$) measured for each SThM probe. When sample roughness increases, the adhesion force decreases by up to 97%, about 30% and 50%, respectively for the Wollaston, Pd and DS probes. This underlines the correlation between the quality of the probe-sample contact and the heat transfer across it. Roughness significantly deteriorates the quality of the probe-sample contact and thus the thermal transfer associated with mechanical contact. The effect seems more pronounced for the Wollaston microprobe than for the nanoprobes. This can be explained by the roughness of the metallic microfilament [15]: the Wollaston probe-planar sample contact is made by a multitude of small contacts, the number of which is decreased when the surface becomes rough, resulting in a significant decrease in the total contact area. Measurements with nanoprobes are less impacted for the studied roughness range.

An upper bound for the curvature radius $R$ at the contact can be obtained by neglecting the influence of the water meniscus on adhesion (note that we don't expect the water meniscus to be predominant for heat transfer [9-15]). $R$ is determined from the adhesion force, measured when retracting the probe from the sample, by using the mechanical model of Rabinovich *et al.* [16-17]. This model considers a rough surface with periodic peak-to-peak distance $L_{RMS}$ and mean out-of-plane deviation $\delta Z_{RMS}$ associated with hemispherical asperities of curvature radius $r = f(\delta Z_{RMS}, L_{RMS})$:

$$F_{ad} = \frac{A_H.R}{6.H_0^2}\frac{1}{\left(1+58.14\frac{R\,\delta Z_{RMS}}{L_{RMS}^2}\right)\left(1+1.817\frac{\delta Z_{RMS}}{H_0}\right)^2} \quad (1)$$

with $A_H \sim 2.65\ 10^{-19}$ J the Hamaker constant and $H_0 = 0.3$ nm the minimum separation distance between the tip apex and the asperity. Using this expression one finds $R = 9 \pm 2$ nm for the DS probe in accordance with previous estimate [12], $R = 6.4 \pm 0.5$ nm for the Pd probe which is ten to five times lower than the values announced by the provider

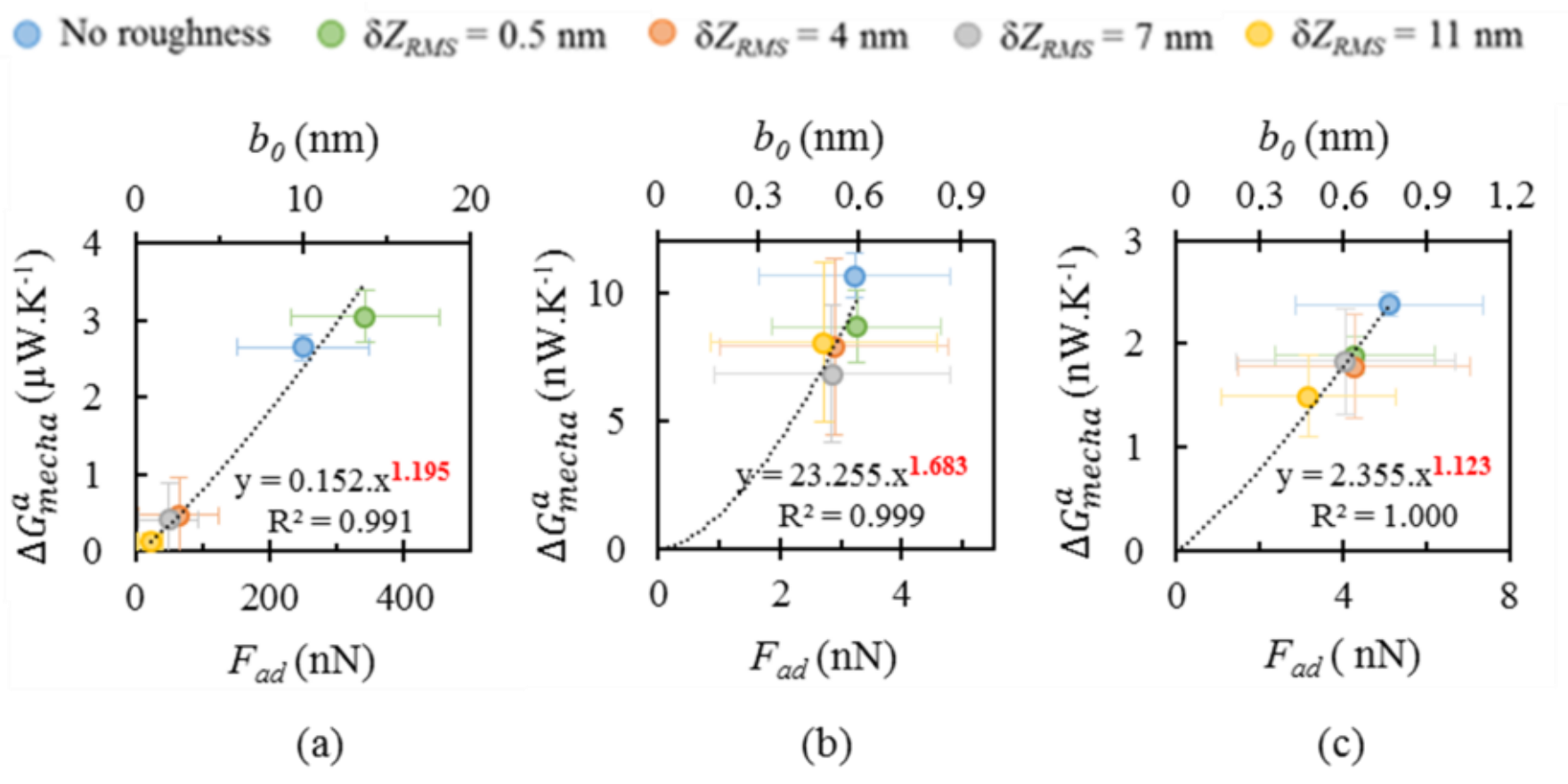

**Fig. 3. Thermal conductance due to mechanical contact as a function of the adhesion force ($F_{ad}$) and corresponding contact radii for experiments performed in air conditions on the rough samples with (a) the Wollaston probe, (b) the Pd probe and (c) the DS probe.**

[12]. This difference could be understood as a contact considered to be made through an asperity at the apex of the tip. For the Wollaston probe that is the largest and roughest probe, $R$ values are very dispersed and the mean is around 450 nm as found in [15]. Eq. (1), which assumes that the surface is rougher than the probe ($R$ > $L_{RMS}$ , $\delta Z_{RMS}$), could therefore be applicable for the Wollaston probe but is only approximate for the two other probes. Adding the Derjaguin-Müller-Toporov (DMT) model for the sphere-plane configuration [18] we can determine, for each surface, a lower bound for the contact radius $b_0$ when zero force is applied:

$$b_0 = \sqrt[3]{\frac{1}{E^*}\frac{R.r}{R+r}F_{ad}} \quad , \qquad (2)$$

where $r$ is the curvature radius of rough surface asperities and $E^*$ is the generalized Young modulus. Fig. 3 provides this quantity for the various probes. For the Wollaston probe $\Delta G^a_{mecha}$ is found proportional to $b_0^{1.2}$, for the Pd nanoprobe to $b_0^{1.6}$and for the DS probe to $b_0^{1.1}$. It is known that when $\Delta G^a_{mecha} \propto b_0$ heat transfer is diffusive, and that $\Delta G^a_{mecha} \propto {b_0}^2$ for a ballistic [19] or thermal-boundary limited (Kapitza) transfer. The oxide layer that is present on the surface of samples and probably on the resistive elements of the probes can be involved in this interfacial thermal resistance. For the three probes the exponent is larger than 1, which suggests that thermal transfer is not only diffusive. Let us note however that $b_0$ values are well below the phonon average mean free path of Si (around 250 nm) so that an exponent closer to 2 would be expected. The role of water meniscus on adhesion, which is here neglected, could explain the difference with such value. A generalization of Eq. (1) to arbitrary values of $R$, which would include contact of the probes' sides with the samples, would be useful to clarify these observations.

It is interesting to analyze the heat transfer reduction in light of the usual SThM measurement process, which involves first a calibration with samples of well-known thermal conductivity with surfaces as flat as possible. Fig. 4 provides such a calibration curve, which underlines the lack of sensitivity for large thermal conductivity. Most importantly, it highlights that using the average value of $\Delta G^a_{mecha}$ for a rough sample of unknown thermal conductivity would lead to determining a thermal conductivity much lower than the correct value (see red arrow), as if an insulating body was present below the surface. Using the upper values of the conductance range may be better (possibly also for the calibration curve) but induces also uncertainty. As a consequence, a detailed analysis of roughness is essential prior to SThM thermal-conductivity determination from the contact.

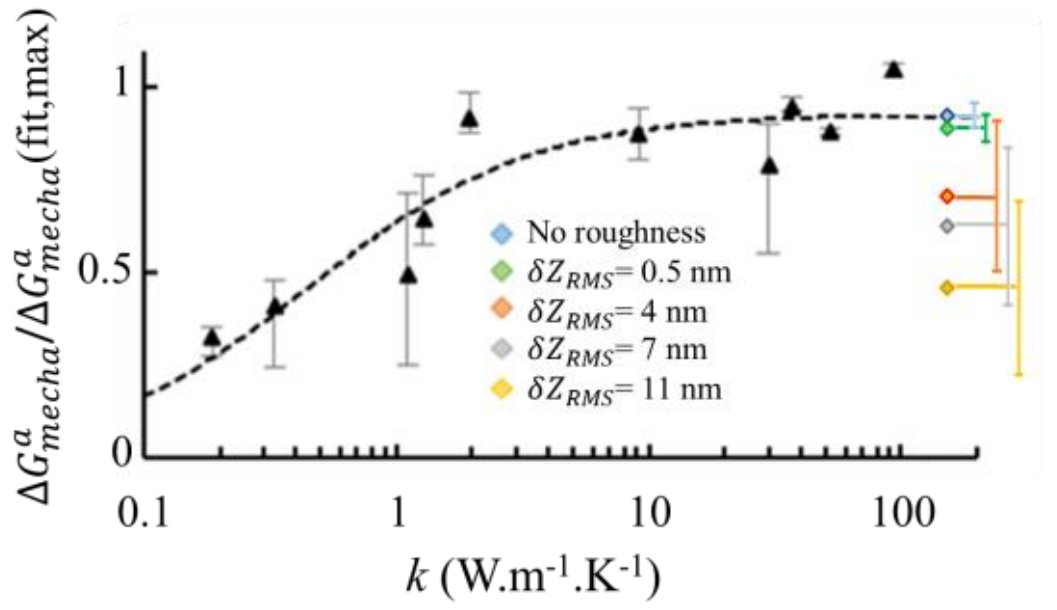

**Fig. 4. Normalized thermal conductance across contact as a function of apparent sample thermal conductivity *k* for both reference and rough samples (Pd probe operated in vacuum). The dashed black line is a calibration curve obtained from average thermal conductances of bulk reference materials with roughness as low as possible (black dots) with a different probe (see Suppl. Mat. for explanations).**

In conclusion, we have measured thermal conductances across micro to nanocontacts by means of SThM probes on Si samples. For roughness $\delta Z_{RMS}$ close to 10 nm, the decrease in contact thermal conductance can reach as much as 90% at a microcontact and about 50% at nanocontacts. In all cases, surface roughness strongly alters the mechanical contact, resulting in most cases in multi-contacts reducing the apparent contact radius. It is found that heat transfer is not only diffusive, but that ballistic or boundary-limited heat conduction can also be involved. Finally, we demonstrate that sample roughness can completely distort the analysis of SThM measurements when estimating thermal conductivity of materials. It will be needed to study the effect of roughness on materials covering the whole thermal conductivity range in order to be able to analyse correctly the thermal data. Another pending issue is that current mechanical models consider usually only the mechanical properties of solid materials, so the water meniscus and its contact radius deserve to be further investigated.

## SUPPLEMENTARY MATERIALS

See supplementary material for further details on SThM probe and sample characterizations, SThM images and protocols.

## ACKNOWLEDGMENTS

The research leading to these results has received funding from the European Union Seventh Framework Programme FP7-NMP-2013-LARGE-7 under grant agreement n°604668.

## DATA AVAILABILITY

The data that support the findings of this study are available from the corresponding author upon reasonable request.

## SUPPLEMENTARY MATERIAL

### A- Probes and their calibrations

Fig. A shows scanning electron microscopy (SEM) images of the three probes. The Wollaston probe (Fig. A(a)) comprises a V-shaped $Pt_{90}/Rd_{10}$ wire of 5 µm in diameter and 200 µm in length. The $Pt_{90}/Rd_{10}$ wire is obtained by removal of the silver shell of the Wollaston wire of diameter 75 µm [A1]. This fabrication process reveals many grooves on the probe surface [A2, A3]. Consequently, it is likely that the first probe-sample contact is established only between one of these grooves and the sample surface. The equivalent curvature radius of an individual groove can be of several hundreds of nanometers [A2]. The spring constant of the probe used was calculated to be $k_r$ = 5 ± 1 N.m$^{-1}$ from the geometrical and physical parameters of the probe [A4]. The overall temperature coefficient of electrical resistance $\alpha = \frac{1}{R}\frac{dR}{dT}$ of the probe was measured about 1.4×10$^{-3}$ K$^{-1}$ from measurements of the electrical resistance of the probe $R$ in an oven as a function of temperature. The electrical resistance $R_p$ of the sensing part of the probe, which is in series with electrical connection, is deduced from dimensional inspection by optical microscopy and scanning electron microscopy. We also verified the dimension of the V-shaped $Pt_{90}/Rd_{10}$ wire from 3ω measurements for this probe [A5]. Its variation can be estimated as a function of the wire mean temperature $\overline{\theta} = \overline{T} - T_a$:

$$\Delta R_P = R_p - R_{p0} = R_{p0}\,\alpha_p\,\overline{\theta} \quad \text{(A1)}$$

where $R_{p0}$ = 2.5 ± 0.4 Ω is the electrical resistance of the sensing part of the wire at room temperature $T_a$ (30 °C) and $\alpha_p$ =1.66×10$^{-3}$ K$^{-1}$ for the $Pt_{90}/Rh_{10}$ wire [A5].

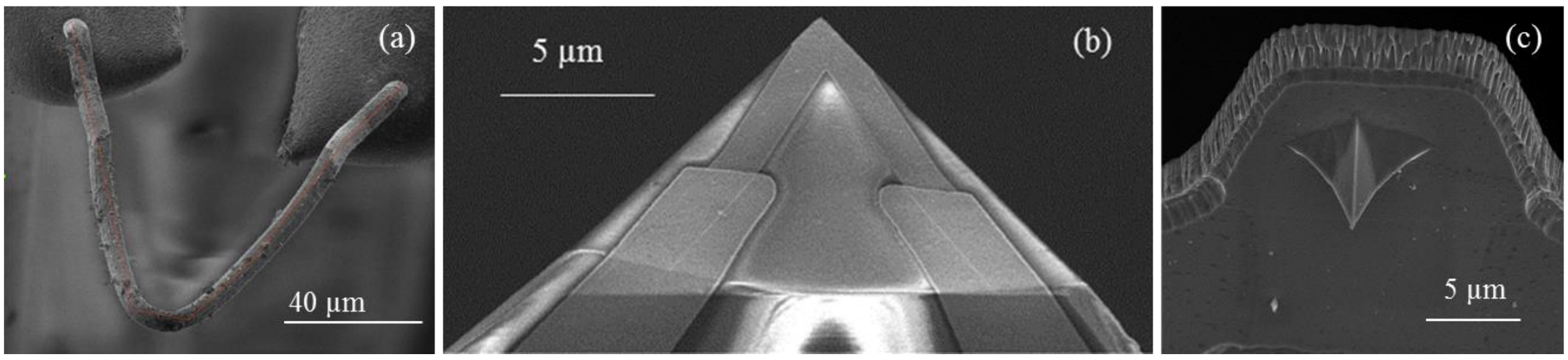

**FIG. A. SEM images of (a) the Wollaston probe, (b) the Pd probe and (c) the DS probe.**

The Pd probe involves a thin resistive Pd film and pads of gold deposited on a silicon nitride ($Si_3N_4$) cantilever (Fig. A(b)). Due to the shape and configuration of the probe apex, the contact with the sample is established through the $Si_3N_4$ part for this probe. The apex curvature radius is smaller than 100 nm [A6, A7] and was estimated of 50 nm using a methodology proposed in references [A2, A8]. The spring constant of the probe was measured of 0.09 ± 0.02 N.m$^{-1}$ using reference lever technique [A9]. The coefficient $\alpha$ of the probe was determined to be about $\alpha$ = 7.8 ± 0.1 × 10$^{-4}$ K$^{-1}$ using the same method as that previously specified for the Wollaston probe. Knowing $\alpha_p$ = 1.37×10$^{-3}$ K$^{-1}$ for the Pd wire and those of the other metallic components of the probe [A8], the variation of the electrical resistance of the Pd film, which is the thermal sensor, can be estimated as a function of the film mean temperature ($R_{p0}$ = 82 ± 8 Ω in Eq. (A1)).

The DS probe consists of two micrometric legs with high doping level and a low-doped resistive element platform. The tip with a curvature radius expected to be close to 10 nm [A7] has an inverted pyramidal shape and is mounted on top of the resistive element (Fig. A(c)). For the characterization of the DS probe we used the same methods as for the other probes. The spring constant of the probe was found to be 0.3 ± 0.1 $N.m^{-1}$. The coefficient $\alpha$ of the probe was determined to be about 2.1 ± 0.2 $10^{-3}$ $K^{-1}$. The electrical resistance of the low-doped resistive element platform was estimated as a function of the probe mean temperature using a quadratic fitting from measurements of the electrical resistance of the probe in an oven at different levels of temperature [A10].

## B- SThM measurement of the contact thermal conductances

After calibration of probes, thermal control units based on balanced Wheatstone bridges were used to monitor the probe mean temperature rise $\bar{\theta}$ with respect to ambient temperature and the electrical power $P_{el}$ dissipated in the probes.

### B-1 Approach curves

For each sample fourteen approach curves were performed and averaged. This was realized for the three probes. Fig. B shows the averaged curve obtained with a DS probe for a pristine Si sample. The variations of the mean probe temperature rise $\bar{\theta}$ with respect to ambient temperature and the power $P_{el}$ with respect to the position far from contact ('out of contact') were used to determine the variation of thermal conductance $\Delta\bar{G}_{global}$ (Fig. B(a)) associated to the total probe-sample heat transfer:

$$\Delta\bar{G}_{global} = \bar{G}_{IC} - \bar{G}_{OC} \ , \qquad \text{(B.1)}$$

where $G_{IC} = P_{el,IC}/\bar{\theta}_{IC}$ and $G_{OC} = P_{el,OC}/\overline{\theta_{OC}}$ are the thermal conductance of the probe-sample system when the probe is in contact (IC index) and out of contact (OC) with the sample, respectively. The total transfer takes place through the surrounding gas, the water meniscus and the mechanical contact. Note that the electrical power varies also with distance due to the probe electrical resistance variation.

The variations $\Delta\bar{\theta}$ and $\Delta P_{el}$ at the probe jump to contact with the sample are used to determine the variation of thermal conductance $\Delta\bar{G}_{mecha}$ associated with the probe-sample heat transfer across the mechanical contact. It can comprise heat transfer through water meniscus. The variation $\Delta\bar{\theta}$ and $\Delta P_{el}$ while the sample indentation increases (Fig. B(b)) are used to study the evolution of $\Delta\bar{G}_{mecha}$ as a function of the probe–sample force. As the mechanical contact occurs with the tip apex we consider in the following rather the tip apex temperature $\theta_{apex}$ than the mean probe temperature $\bar{\theta}$. In this condition

$$\Delta G^{a}_{mecha} \ = 1/K.\Delta\bar{G}_{mecha} \qquad \text{(B.2)}$$

with $K$a coefficient that is $K = \frac{3}{2}$ for the Wollaston probe [A11], $K = 1.42$ for the Pd probe and $K = 1.08 - 1.23$ for the DS probe. These coefficients were obtained from modelling including full FEM simulations of the temperature profiles in the probes with heuristic parameters when necessary [A12].

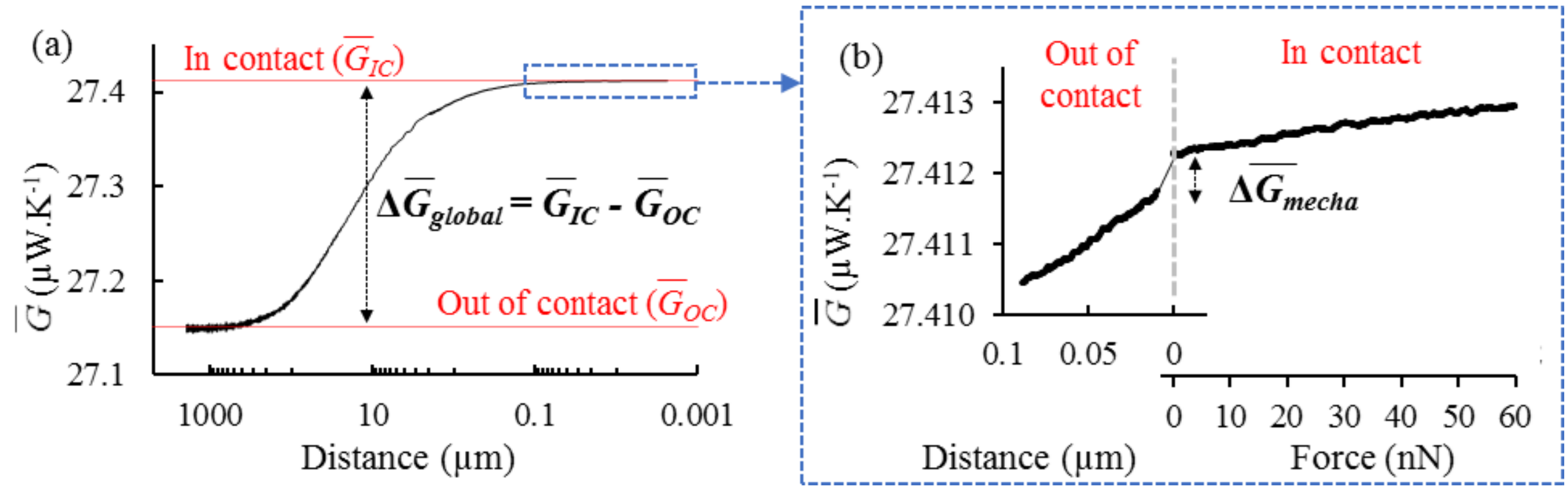

**FIG. B. Thermal conductance of the probe-sample system $\overline{G}$ as a function of distance to contact with a flat Si reference sample (DS probe in air). (a) Thermal conductances $\Delta\overline{G}_{global}$ including all heat transfer channels and (b) $\Delta\overline{G}_{mecha}$ at the contact jump and as a function of the probe-sample force.**

Between measurements of each rough sample two reference samples of silica and silicon (with roughness parameter $\delta Z_{RMS}$ lower than 0.1 nm) were measured to check any change of the probe apex (contamination or damage). No probe change was observed during the experiments performed for this study.

To assess the impact of surface roughness on the SThM measurement, approach curves and images were made both in the ambient air (ambient temperature $T_a \simeq 30$ °C) and in primary vacuum (pressure $P = 0.28$ mbar and $T_a \simeq 30$ °C). The probes were heated at $\bar{\theta}_{OC} = \bar{T}_{p,OC} - T_a = 70$ K for the Wollaston probe, $\bar{\theta}_{OC} = 65$ K for the Pd probe and $\bar{\theta}_{OC} = 90$ K for the DS probe.

## B-2 Image measurements

For each sample thermal images of various area (1 µm², 4 µm², 9 µm², 25 µm² and 100 µm²) were performed with the three probes in air and vacuum conditions (Fig. C). The force applied between the tip and the sample was constant during a scan. It was set to about 500 nN for Wollaston probe, 10 nN for Pd probe and 70 nN for the DS probe. Particular attention was paid to the acquisition time of each image pixel so that it corresponds to about 5 times the thermal response time of the probe. The number and size of pixels per image vary.

The thermal signal studied corresponds to $\Delta G^a_{mecha}$. For each thermal image the average value $\overline{\Delta G^a_{mecha}}$ and its rms dispersion $\sigma$ are reported as a function of the Si surface roughness in Fig. D. These results show, as with point-measurements, a marked decrease in thermal conductance with the increase in the sample roughness. It is observed that in these experimental conditions, i.e. when working with constant force and contact is well established, the decrease in roughness $\overline{\Delta G^a_{mecha}}$ varies linearly. This result is different from what was observed during the first probe-sample contact during point measurements: the decrease in roughness $\overline{\Delta G^a_{mecha}}$ varies much more rapidly (Fig. 2 in the core manuscript) in point measurements. However, the total decrease observed is similar, or slightly greater, than that estimated from point measurements: up to 90% for the Wollaston probe, 40% for the Pd probe and 60% for the DS probe. It also appears that the dispersion $\sigma$ increases with the roughness, which induces a more variable mechanical contact during the scan. These results are equivalent regardless of the size of the image. The conclusions are therefore similar to those obtained during the point measurements.

| | **Wollaston probe** | **Pd probe** | **DS probe** |
|---|---|---|---|
| Non-rough sample | 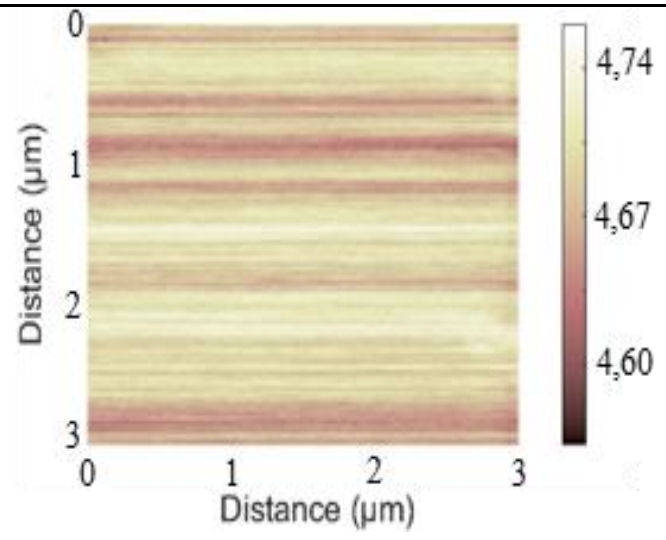 $\overline{\Delta G^{a}_{mecha}} = 4.685$ μW. K$^{-1}$ <br> $\sigma = 0.035$ μW.K$^{-1}$ | 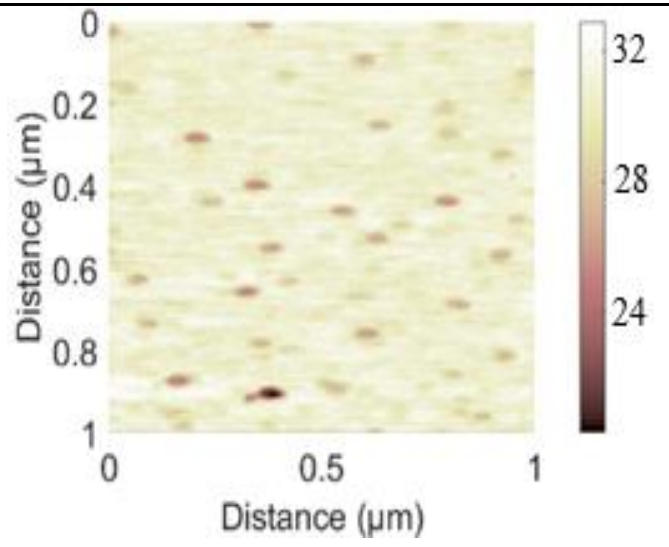 $\overline{\Delta G^{a}_{mecha}} = 30.46$ nW.K$^{-1}$ <br> $\sigma = 1.12$ nW.K$^{-1}$ | 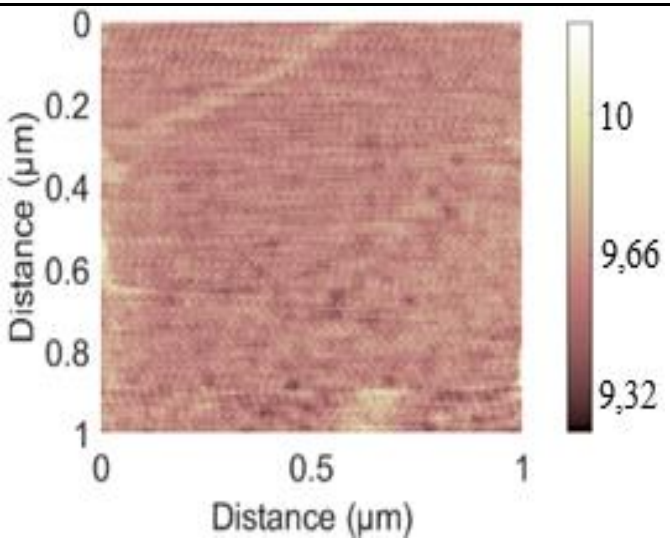 $\overline{\Delta G^{a}_{mecha}} = 9.63$ nW.K$^{-1}$ <br> $\sigma = 0.13$ nW.K$^{-1}$ |
| $\delta Z_{RMS}$ = 0.5 nm | 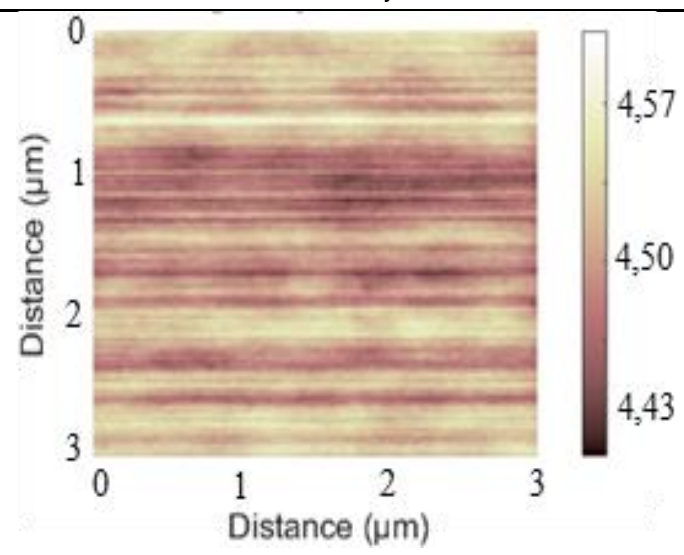 $\overline{\Delta G^{a}_{mecha}} = 4.508$ μW. K$^{-1}$ <br> $\sigma = 0.042$ μW.K$^{-1}$ | 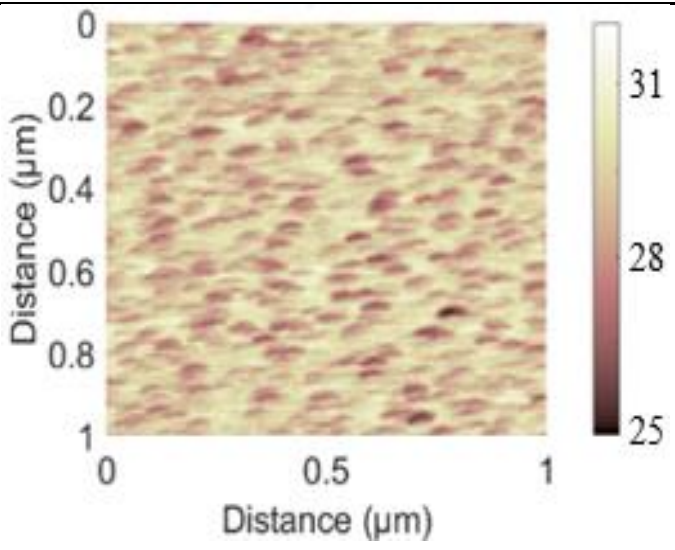 $\overline{\Delta G^{a}_{mecha}} = 29.30$ nW.K$^{-1}$ <br> $\sigma = 1.20$ nW.K$^{-1}$ | 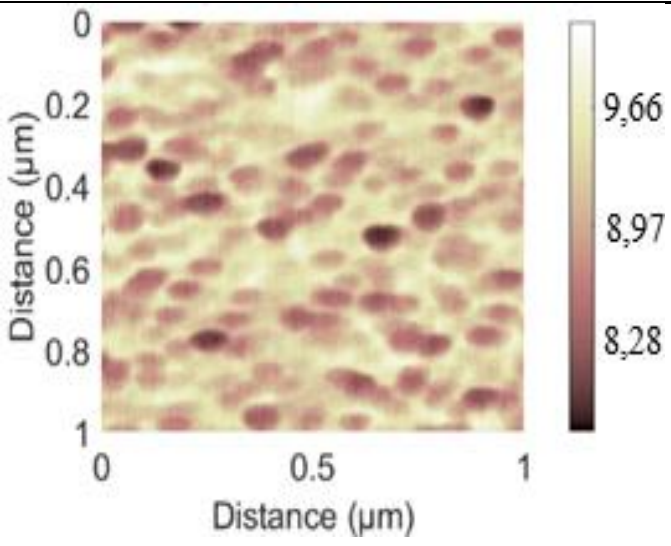 $\overline{\Delta G^{a}_{mecha}} = 9.16$ nW.K$^{-1}$ <br> $\sigma = 0.53$ nW.K$^{-1}$ |
| $\delta Z_{RMS}$ = 4 nm | 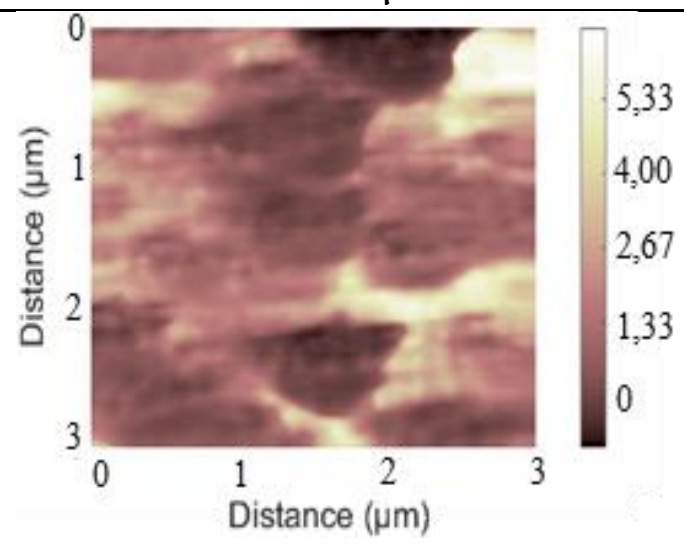 $\overline{\Delta G^{a}_{mecha}} = 1.497$ μW. K$^{-1}$ <br> $\sigma = 2.152$ μW.K$^{-1}$ | 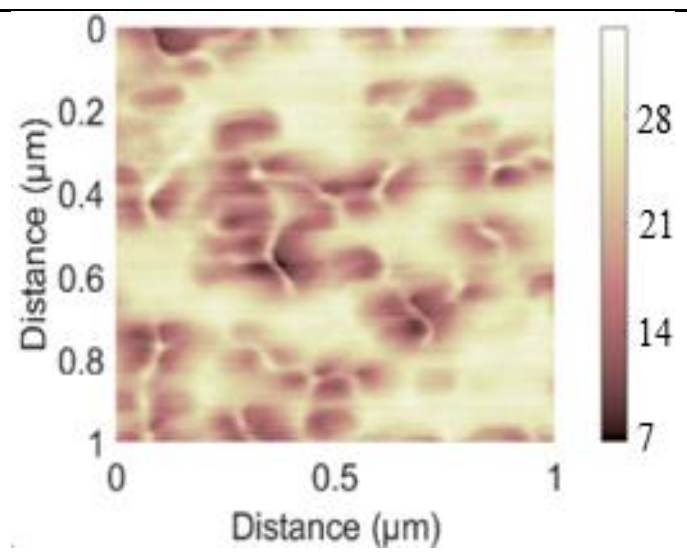 $\overline{\Delta G^{a}_{mecha}} = 23.29$ nW.K$^{-1}$ <br> $\sigma = 6.77$ nW.K$^{-1}$ | 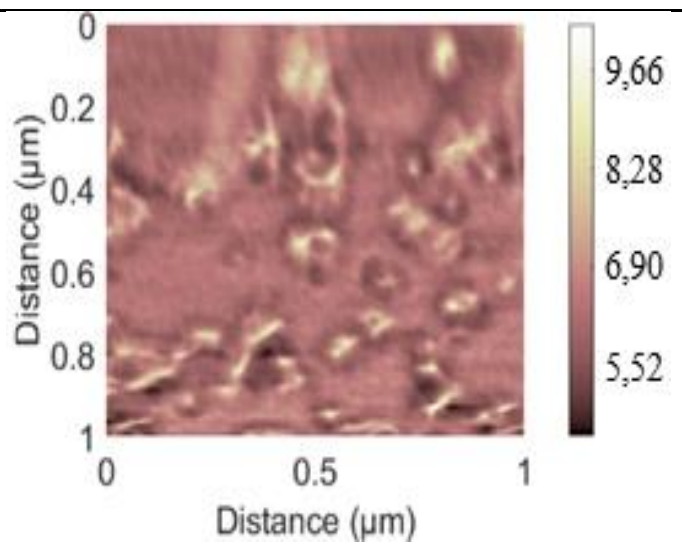 $\overline{\Delta G^{a}_{mecha}} = 6.33$ nW.K$^{-1}$ <br> $\sigma = 0.84$ nW.K$^{-1}$ |
| $\delta Z_{RMS}$ = 7 nm | 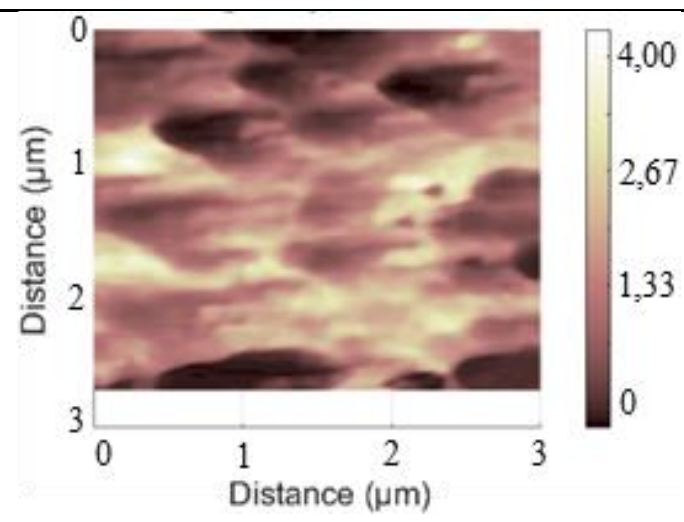 $\overline{\Delta G^{a}_{mecha}} = 1.245$ μW. K$^{-1}$ <br> $\sigma = 1.405$ μW.K$^{-1}$ | 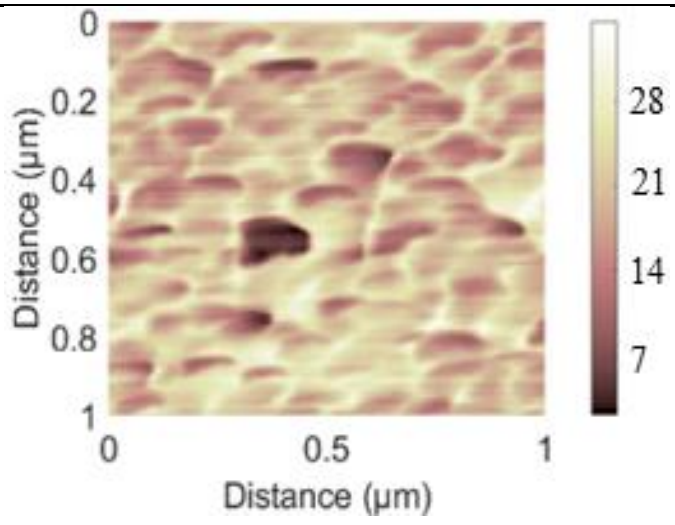 $\overline{\Delta G^{a}_{mecha}} = 20.60$ nW.K$^{-1}$ <br> $\sigma = 7.08$ nW.K$^{-1}$ | 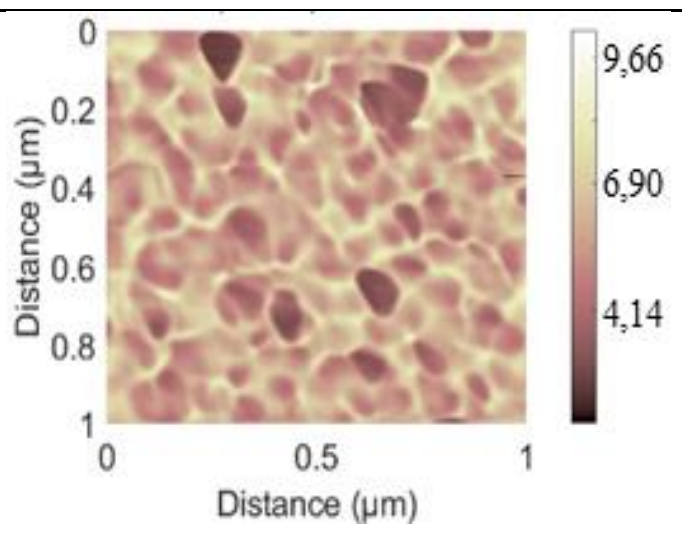 $\overline{\Delta G^{a}_{mecha}} = 5.91$ nW.K$^{-1}$ <br> $\sigma = 1.62$ nW.K$^{-1}$ |
| $\delta Z_{RMS}$ = 11 nm | 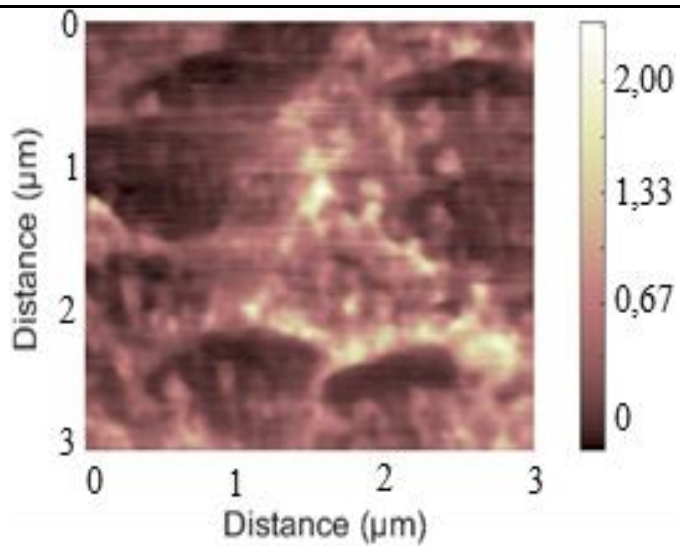 $\overline{\Delta G^{a}_{mecha}} = 0.346$ μW. K$^{-1}$ <br> $\sigma = 0.571$ μW.K$^{-1}$ | 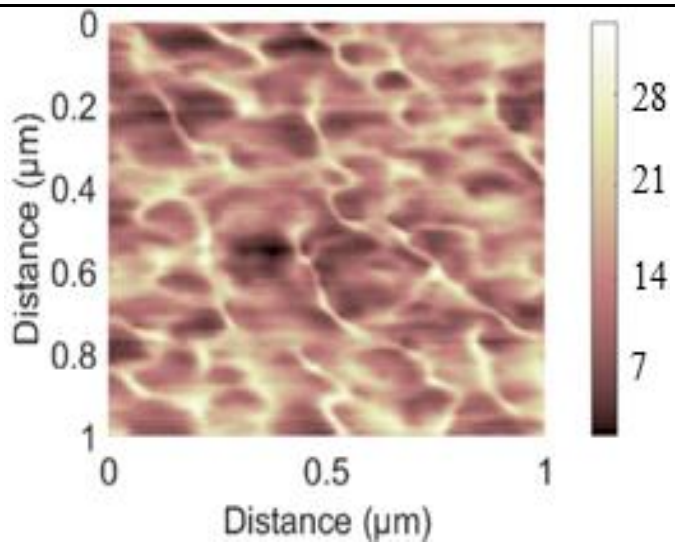 $\overline{\Delta G^{a}_{mecha}} = 15.10$ nW.K$^{-1}$ <br> $\sigma = 7.69$ nW.K$^{-1}$ | 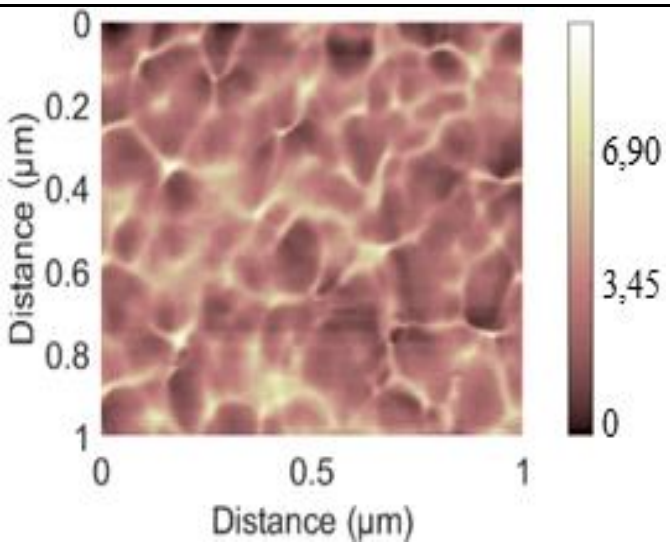 $\overline{\Delta G^{a}_{mecha}} = 3.33$ nW.K$^{-1}$ <br> $\sigma = 2.04$ nW.K$^{-1}$ |

**FIG. C. Images of thermal conductance $\Delta G^{a}_{mecha}$ obtained in vacuum conditions with the Wollaston probe for a scanned surface of 3x3 µm², and for the Pd and the DS probes for a scanned surface of 1x1 µm². The average $\overline{\Delta G^{a}_{mecha}}$ and rms dispersion $\sigma$ are indicated.**

We observe in Fig. D that for image measurements, performed at constant applied force, the values of $\overline{\Delta G^{a}_{mecha}}$ measured in vacuum are systematically larger than those measured under ambient air for the pristine surface and the slightly-rougher surface ($\delta Z_{RMS}$ = 0.5 nm). This difference does not appear as strong for point measurements (Fig. 2) when scanning, i.e. at constant applied force, than when measuring on selected points, i.e. for the force at first contact. The absolute difference weakens when roughness increases. This can be explained by noticing that keeping the same applied force in ambient and in vacuum does not at all guarantee that the mechanical contact area stays identical. In vacuum, the meniscus volume is expected to be smaller than that in ambient conditions [A13]. So the share of the applied force associated with the mechanical contact between the solids is larger, which means that the solid contact area is larger than in air. Assuming that the main part of the flux is transferred at the mechanical contact and not in the meniscus [A2, A8], this analysis means that $\overline{\Delta G^{a}_{mecha}}$ is larger in vacuum than in air. This is exactly what is observed. To have more information about the effect of the meniscus, it would be interesting to carry out the experiments as a function of temperature in order to modify the meniscus shape [A2, A8].

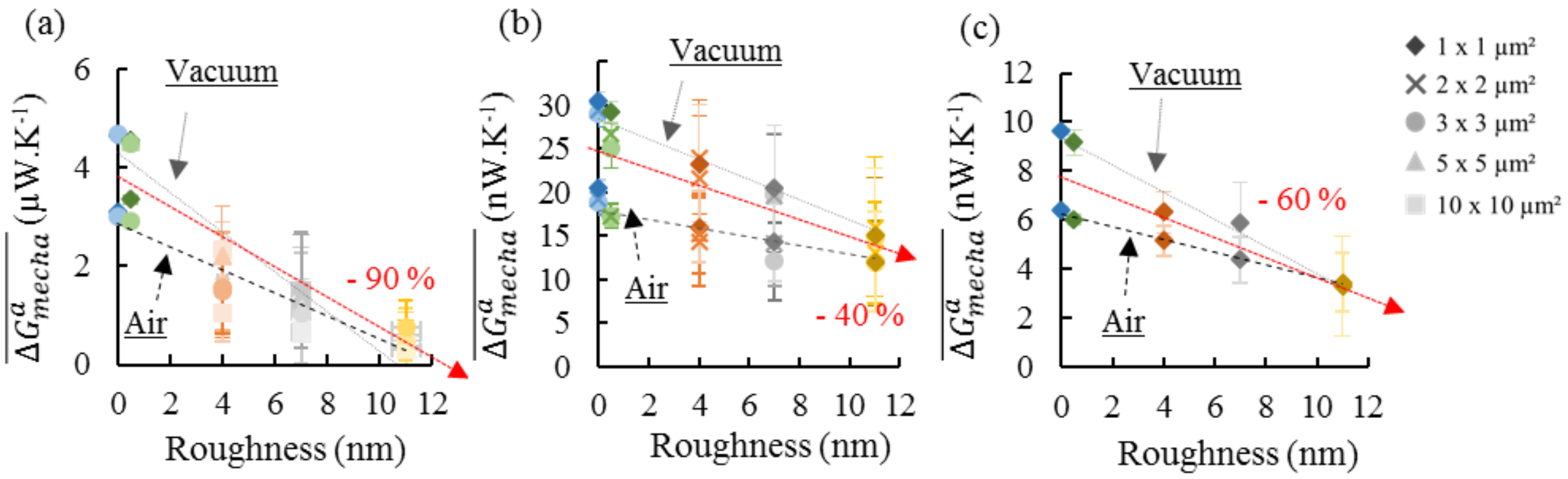

**Fig. D. Thermal conductance $\overline{\Delta G^{a}_{mecha}}$ estimated from images of rough Si samples with different areas (ambient and vacuum conditions) with the (a) Wollaston, (b) Pd and (c) DS probes.**

## C- Discussion

### Impact of roughness on thermal conductivity measurements using SThM

Fig 4 in the core manuscript reports on the measurements of the thermal conductance across mechanical contact measured with the Pd probe under vacuum conditions on reference bulk samples of well-known thermal conductivity $k$ and known out-of-plane roughness $\delta Z_{RMS}$. The calibration curve was plotted with reference bulk material thermal conductivities determined from values of thermal diffusivity, specific heat and density measured respectively by laser flash method, differential scanning calorimetry (DSC) and Archimedean immersion method by the French NMI LNE. The obtained values are given in Table A. Roughness was measured using AFM.

The dashed black line in Fig. 4 of the core manuscript was obtained from a simple fit with the following function:

$$\Delta G^{a}_{mecha} = \frac{A}{1+\frac{B}{k}} \quad . \tag{C.1}$$

It has been popular to identify the parameters ($A$,$B$) as follows: $A = G_{c-mecha}$ and $B = \frac{G_{c-mecha}}{4.b_{mecha}}$. This identification stems from a model of thermal resistances where $1/G_{c-mecha}$ and $1/G_{sample}$ are connected in series. However, the identification involves the strong assumptions that the interfacial conductance $G_{c-mecha}$ and the mechanical contact radius $b_{mecha}$ are independent of sample physical properties and that the thermal spreading resistance of the sample is diffusive, which are at least uncertain [A14]. Other more-advanced models would provide different curves, maybe closer to the measured values, but always with the same trend and probably at the cost of requiring more fit parameters. Here, Eq. (C.1) is considered without the identification, as a simple fit function with a low number of parameters.

It is important to note that the calibration curve was performed with another probe than that used for the roughness measurements (two different experiments). As a result, there is a difference between the probe response curves due to natural variation from probe to probe. In Fig. 4 of the core manuscript, we have decided to normalize $\Delta G^{a}_{mecha}$ by the maximal values expected. Here below (Fig. E), we show the raw data which highlight the difference in scales. This further underlines the difficulty of uncalibrated SThM measurements.

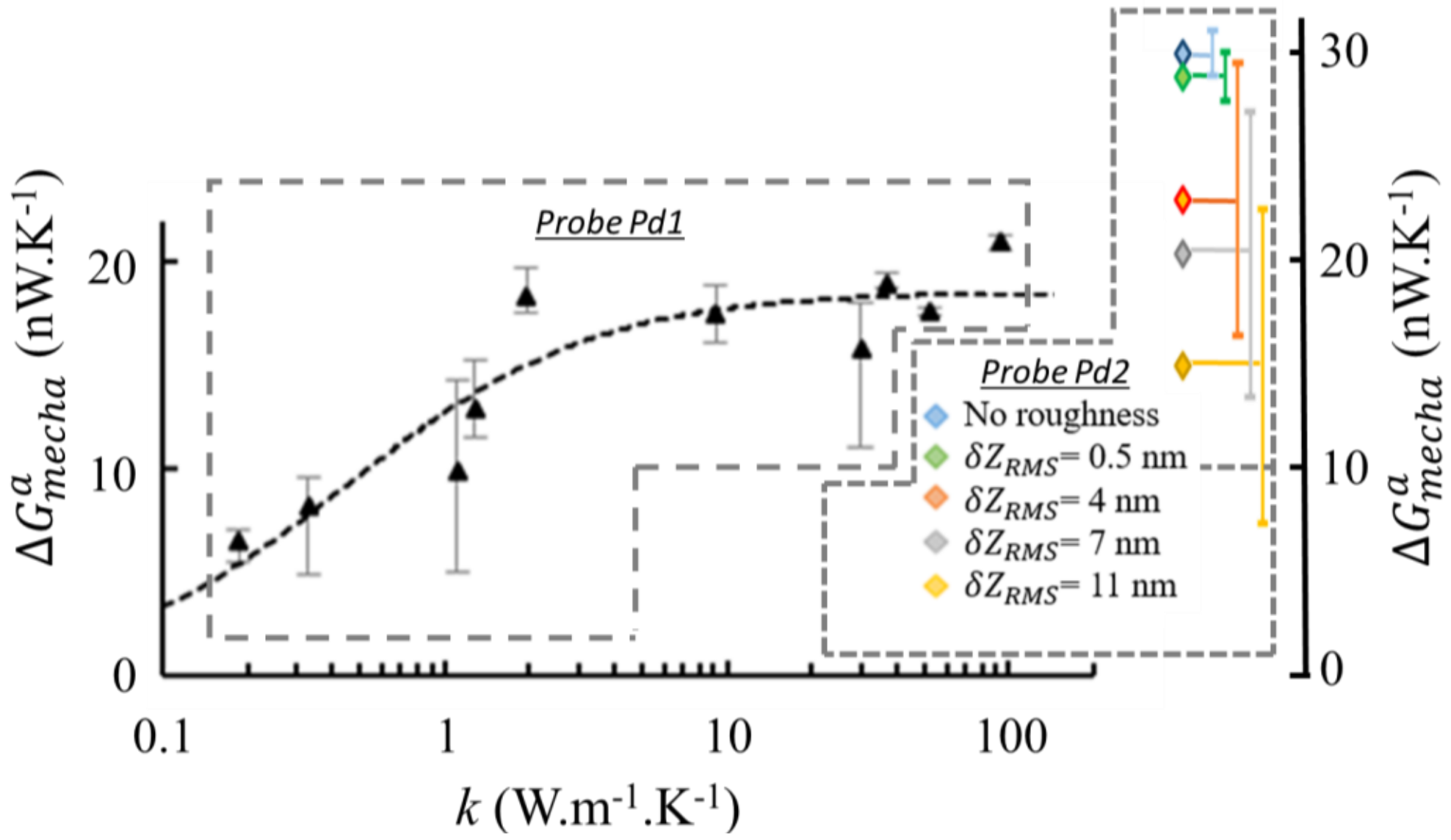

**Fig. E. Thermal conductance across contact as a function of apparent sample thermal conductivity *k* for both reference and rough samples (Pd probe operated in vacuum). The reference and rough samples were measured with different probes.**

Finally, let us comment on the impact of roughness on samples of lower thermal conductivity than Si, in the range where SThM is more sensitive ($k$ lower than 3 W.m$^{-1}$.K$^{-1}$). Heat dissipation should be fully diffusive for such materials of shorter mean free paths and where the volume thermal conductance is larger than that at the surface, reducing the impact of surface roughness. For such materials, the dependence of thermal conductance on the mechanical contact radius could therefore be smaller than for high-thermal conductivity materials. The impact of roughness would therefore be weaker than for Si. Detailed studies of the roughness impact on measurement for insulating materials would be useful to confirm this analysis.

**Table A. List and properties of reference bulk materials.**

| Material | Thermal conductivity $k$ (W $m^{-1}K^{-1}$) | Roughness $\delta Z_{RMS}$ (nm) |
|---|---|---|
| PMMA | 0.187 | 6.43 |
| POM-C | 0.329 | 15.7 |
| Glass | 1.11 | 1.90 |
| Fused $SiO_2$ | 1.28 | 0.73 |
| $ZrO_2$ | 1.95 | 1.58 |
| $Al_2O_3$ | 29.8 | 10.0 |
| Single crystal $Al_2O_3$ | 36.9 | 0.69 |
| Ge | 52.0 | 0.57 |
| Si n++ | 71,2 | 0,73 |
| Si p++ | 93.4 | 1.47 |

## References for the supplementary material

Note that in air or when in contact the parabolic profile in the Wollaston probe is bent into an hyperbolic one, but slightly only, so this coefficient is only weakly modified.